# ON THE ON THE CLASSICAL-QUANTUM TRANSITION IN HEAT PHENOMENA INDUCED BY CONTINOUS HIGH ENERGY LASER PULSES

Janina Marciak-Kozlowska[1] and Miroslaw Kozlowski[*]

[1]Institute of Electron Technology,

Al. Lotników 32/46, 02-668 Warsaw, Poland

________________________________________________________

[*]Corresponding author, e-mail: MiroslawKozlowski@aster.pl

Abstract

In this paper the continuous laser beam interaction with matter is investigated. The velocities of the thermal propagation are calculated. It is shown that for the value of the product $\omega\tau > 1$, $\omega$ is the angular frequency of the laser beam and $\tau$ is the relaxation time for thermal phenomena the Heisenberg uncertainty $VT_P > \dfrac{h}{2}$ is fulfilled, $V$ is the potential energy of the heat carriers and $T_P$ is the period of the laser beam.

## 1. INTRODUCTION

Recently the emergent new technologies, especially in nanoelectronics have led to increasing interest in quantum effects in technological application. One of that effect is the decoherence – the processes that limit the appearance of quantum effects and turn them into classical phenomena [1].

In this paper we study the quantum thermal phenomena. When continuous laser beam with angular frequency ω interacts with matter the hyperbolic thermal diffusion can be observed [2]. In the case of high energy beam the natural frame work to description of quantum thermal phenomena is the quantum hyperbolic transport equation [2].

The most stringent restriction for quantum phenomena is the Heisenberg uncertainty relation. In this paper we obtain Heisenberg uncertainty for thermal phenomena. We study the velocity of the heat propagation and find out it depend strongly on the product of ωτ where $\tau$ is the relaxation time for thermal phenomena. As the result we obtain that for $\omega\tau > 1$ the potential energy $V$ for heat carriers and period of the laser beam $T_P$ fulfill the Heisenberg uncertainty relation

$$VT_P > \frac{h}{2}.$$

The result of the paper offers the new insight in the transition from classical to quantum level of the description of the transport phenomena.

## 2. THE MODEL EQUATIONS

In monograph [2] the hyperbolic diffusion equation was discussed, viz.:

$$\tau\frac{\partial^2 T}{\partial t^2} + \frac{\partial T}{\partial t} = D\frac{\partial^2 T}{\partial x^2}. \qquad (1)$$

In Eq. (1), $\tau$ denotes the relaxation time for thermal phenomena, $D$ is the thermal diffusion coefficient. For very short relaxation time, $\tau \to 0$ Eq. (1) is the Fourier equation for heat diffusion

$$\frac{\partial T}{\partial t} = D\frac{\partial^2 T}{\partial t^2}. \qquad (2)$$

Recently the hyperbolic diffusion equation (1) for materials with non-homogenous inner structure has run into a serious controversy. While one group of investigators has observed very strong evidence of hyperbolic nature of conduction in such materials and experimentally determined the corresponding relaxation times to be of the order of tens of seconds [2], the other group proclaims that their experiments do not show any such hyperbolic behaviour and the classical Fourier law is good enough to describe thermal transport in them [3, 4]. Very recently W. Roetzel et al. [5] confirmed a hyperbolic behaviour of thermal propagation.

In this paper we investigate the velocity of thermal phenomena induced by laser pulses. First of all we discuss the classical diffusion equation, Fourier law (2). We will seek the solution of equation (2) in the form

$$T(x,t) = f(x)e^{i\omega t}. \tag{3}$$

After substituting Eq. (3) to Eq. (2) we obtain the new equation

$$\frac{d^2 f(x)}{dx^2} = \frac{i\omega}{D} f(x). \tag{4}$$

Let as assume the solution of Eq. (4) as exponential function

$$f(x) = e^{qx}. \tag{5}$$

In that case we obtain for $f(x)$, viz.

$$f(x) = e^{-\left(\frac{\omega}{2D}\right)^{1/2}(1+i)} \tag{6}$$

and

$$T(x,t) = e^{-\left(\frac{\omega}{2D}\right)^{1/2}} e^{i\left(\omega t - \left(\frac{\omega}{2D}\right)^{1/2} x\right)}. \tag{7}$$

We define the diffusion velocity from Eq. (7) viz.:

$$\omega - \left(\frac{\omega}{2D}\right)^{\frac{1}{2}} \frac{dx}{dt} = 0$$

i.e.

$$v = \frac{dx}{dt} = (2D\omega)^{\frac{1}{2}}. \tag{8}$$

From Eq. (8) we conclude that $v \to \infty$ when $\omega \to \infty$, i.e. the parabolic diffusion, Fourier law (2) is in opposition to special relativity theory, for $v > c$, $c$ is light velocity in vacuum.

For hyperbolic diffusion, Eq. (1), we seek the solution in the form

$$T(x,t) = f(x)e^{i\omega t} \tag{9}$$

After substituting Eq. (9) to Eq. (1) one obtains the new equation

$$D\frac{d^2 f(x)}{dt^2} = i\omega f(x) - \tau\omega^2 f(x) \tag{10}$$

We seek solution of Eq. (10) in the form

$$f(x) = e^{qx} \tag{11}$$

and after the minor algebra one obtains

$$q = -\left(-\frac{\tau\omega^2}{2D} + \frac{\omega}{2D}\left(1+\tau^2\omega^2\right)^{\frac{1}{2}}\right)^{\frac{1}{2}} - \frac{i\dfrac{\omega}{2D}}{\left(-\dfrac{\tau\omega^2}{2D} + \dfrac{\omega}{2D}\left(1+\tau^2\omega^2\right)^{\frac{1}{2}}\right)}. \qquad (12)$$

Considering formulae (9), (11) and (12) we obtain the solution of hyperbolic diffusion equation as, viz.:

$$T(x,t) = e^{-\left(-\frac{\tau\omega^2}{2D}+\frac{\omega}{2D}(1+\tau^2\omega^2)^{\frac{1}{2}}\right)^{\frac{1}{2}}} e^{i\left(\omega t - \frac{\frac{\omega}{2D}}{\left(-\frac{\tau\omega^2}{2D}+\frac{\omega}{2D}(1+\tau^2\omega^2)^{1/2}\right)^{1/2}}x\right)}. \qquad (13)$$

From formula (13) we obtain the velocity of hyperbolic diffusion

$$v = \frac{dx}{dt} = 2D\left(-\frac{\tau\omega^2}{2D} + \frac{\omega}{2D}\left(1+\tau^2\omega^2\right)^{\frac{1}{2}}\right)^{\frac{1}{2}}. \qquad (14)$$

From Eq. (14) we conclude that for $\omega \to \infty$, $v \to 0$.

In the quantum limit of heat transport Eq. (1) has the form

$$\tau\frac{\partial^2 T}{\partial t^2} + \frac{\partial T}{\partial t} - \frac{2V}{\hbar}T = \frac{\hbar}{m}\frac{\partial^2 T}{\partial x^2}. \qquad (15)$$

In Eq. (15) $\tau$ is the relaxation time, $\hbar$ denotes the Planck constant, $V$ is the potential energy and $m$ denotes the mass of heat carriers. We will seek the solution of Eq. (15) in the form

$$T(x,t) = f(x)e^{i\omega t} \qquad (16)$$

where $\omega$ is the angular frequency of the laser light. After substituting Eq. (16) to Eq. (15) one obtains

$$-\tau\omega^2 f(x) + i\omega f(x) + \frac{2V}{\hbar}f(x) - \frac{\hbar}{m}\frac{d^2 f(x)}{dx^2} = 0. \qquad (17)$$

Assuming for $f(x)$, viz.:

$$f(x) = e^{qx}. \qquad (18)$$

We obtain the equation for parameter $q$

$$q^2 = \frac{m}{\hbar}\left(\frac{2V}{\hbar} - \tau\omega^2 + i\omega\right). \qquad (19)$$

The solution of Eq. (19) is

$$q = -\sqrt{\frac{m}{2\hbar}\left(\frac{2V}{\hbar} - \tau\omega^2\right) + \frac{m}{2\hbar}\sqrt{\omega^2 + \left(\frac{2V}{\hbar} - \tau\omega^2\right)^2}} -$$
$$i\frac{m\omega}{2\hbar\sqrt{\frac{m}{2\hbar}\left(\frac{2V}{\hbar} - \tau\omega^2\right) + \frac{m}{2\hbar}\sqrt{\omega^2 + \left(\frac{2V}{\hbar} - \tau\omega^2\right)^2}}} \tag{20}$$

and the temperature field $T(x, t)$ — solution of Eq. (1) has the form

$$T(x,t) = e^{-x\sqrt{\frac{m}{2\hbar}\left(\frac{2V}{\hbar} - \tau\omega^2\right)^2 + \frac{m}{2\hbar}\sqrt{\omega^2 + \left(\frac{2V}{\hbar} - \tau\omega^2\right)^2}}} \times$$
$$e^{i\left(\omega t - \frac{m\omega x}{2\hbar\sqrt{\frac{m}{2\hbar}\left(\frac{2V}{\hbar} - \tau\omega^2\right) + \frac{m}{2\hbar}\sqrt{\omega^2 + \left(\frac{2V}{\hbar} - \tau\omega^2\right)^2}}}\right)} \tag{21}$$

From formula (21) we calculate the velocity of heat propagation

$$v = \frac{2\hbar}{m}\sqrt{\frac{m}{2\hbar}\left(\frac{2V}{\hbar} - \tau\omega^2\right) + \frac{m}{2\hbar}\sqrt{\omega^2 + \left(\frac{2V}{\hbar} - \tau\omega^2\right)^2}}. \tag{22}$$

For $V \to 0$, we obtain from formula (22) viz.:

$$v = \frac{2\hbar}{m}\sqrt{-\frac{m}{2\hbar}\tau\omega^2 + \frac{m\omega}{2\hbar}\left(1 + \tau^2\omega^2\right)^{\frac{1}{2}}}. \tag{23}$$

Comparing formulae (14) and (23) we conclude that diffusion coefficient $D$ is equal

$$D = \frac{2\hbar}{m} \tag{24}$$

for quantum limit of heat transport. The most interesting result of the formulae (21) and (22) is that for the relation

$$\frac{2V}{\hbar} - \tau\omega^2 = 0 \tag{25}$$

we conclude that velocity of heat propagation $v$, formula (23) is equals

$$v = (2D\omega)^{\frac{1}{2}}. \tag{26}$$

In Figs. 1(a), 1(b), the relative thermal velocity $v/c$ as the function of $\omega$ and $V$, for $\tau = 0$, i.e. for the Fourier law [2]. From Figs. 1(a) and 1(b) we conclude that for $\omega > 10^{-3}$ c/fm i.e. for $\omega > 10^{20}$ s$^{-1}$, $v/c > 1$ in the opposition to the special relativity.

In Fig. 2(a), 2(b) the relative thermal velocity $v/c$ is calculated for $\tau = 10^{-17}$ s. In that case $v/c < 1$ for all the investigated omega region and is constant.

## 3. THE HEISENBERG UNCERTAINTY PRINCIPLE FOR THERMAL PROCESSES

The condition (25) was first described for O. Heaviside in the context of his investigation of pulse propagation in the transmission time. When we rewrite the formula (25) as

$$\frac{2V}{\omega} = \tau \omega \hbar \qquad (27)$$

or

$$VT_P = \frac{\tau \omega h}{2} \qquad (28)$$

where $T_P$ is the period of temperature oscillations, we conclude that

$$VT_P < \frac{h}{2} \qquad \text{for} \qquad \omega\tau < 1, \qquad (29)$$

$$VT_P > \frac{h}{2} \qquad \text{for} \qquad \omega\tau > 1. \qquad (30)$$

From formulae (29) and (30) we conclude that with increasing the value of $\omega\tau$, we cross the border between classical for $\omega\tau < 1$ and quantum description for $\omega\tau > 1$. In fact the formula (30) is the quantum Heisenberg uncertainty principle.

CONCLUSION

In this paper the Heisenberg uncertainty relation for quantum heat propagation is obtained. It is shown that for continuous laser beam with angular frequency $\omega$, when $\omega\tau > 1$ the Heisenberg inequality holds this is not the case for heat propagation induced by laser beam with $\omega\tau < 1$.

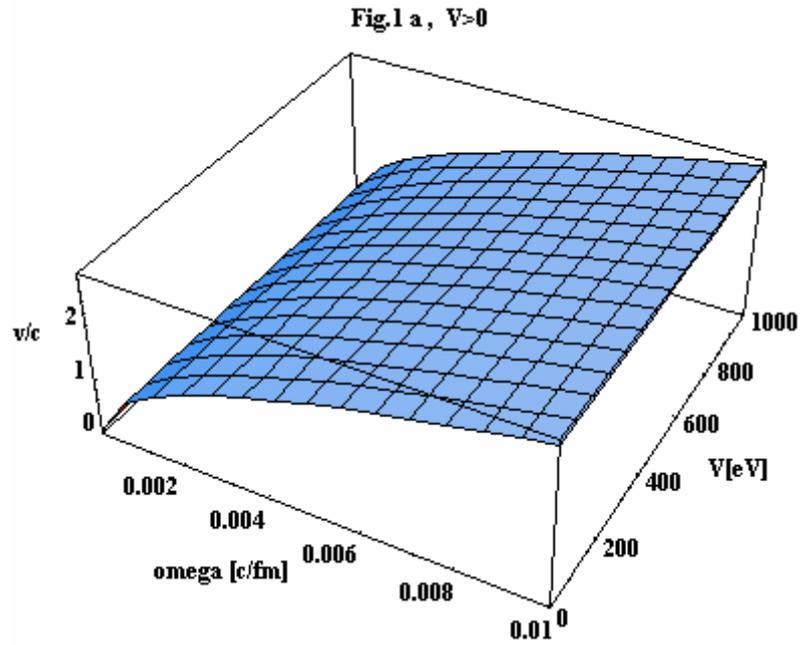

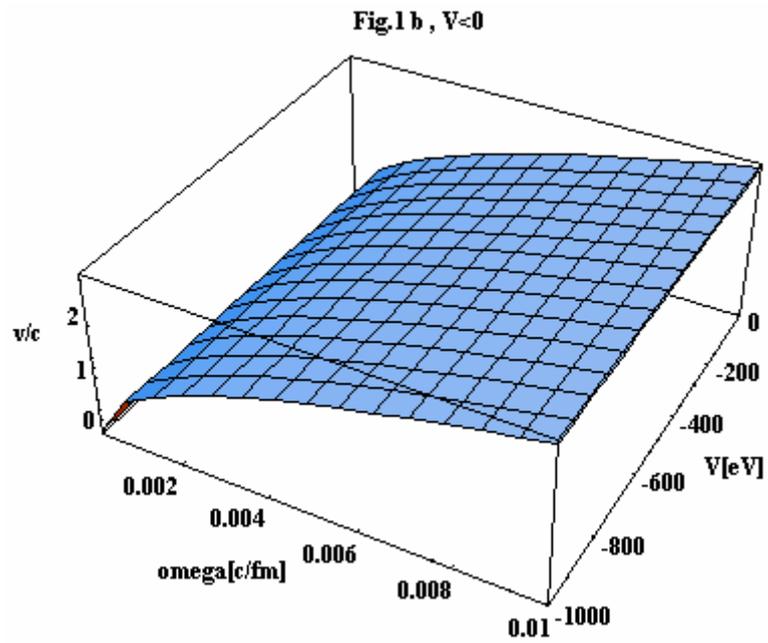

Fig. 1(a) The relative speed of heat propagation for the Fourier approximation $\tau = 0$, $V > 0$, (b) The same for $V < 0$.

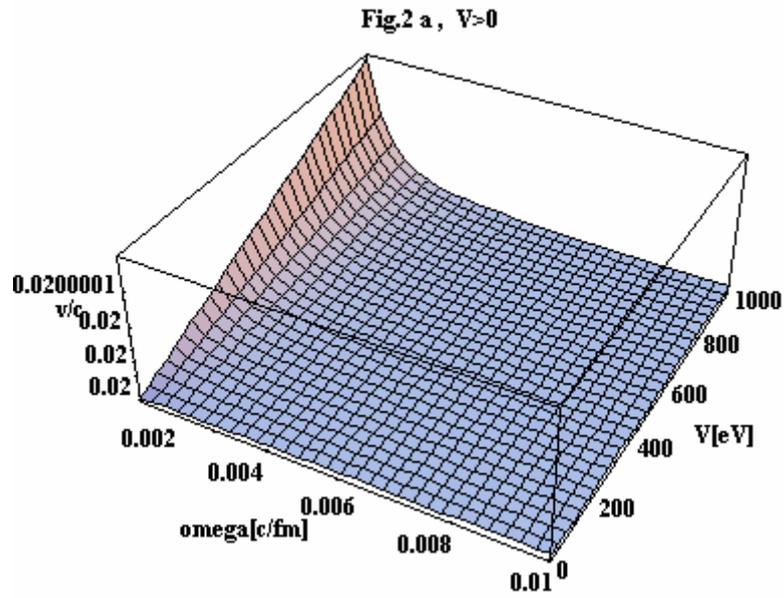

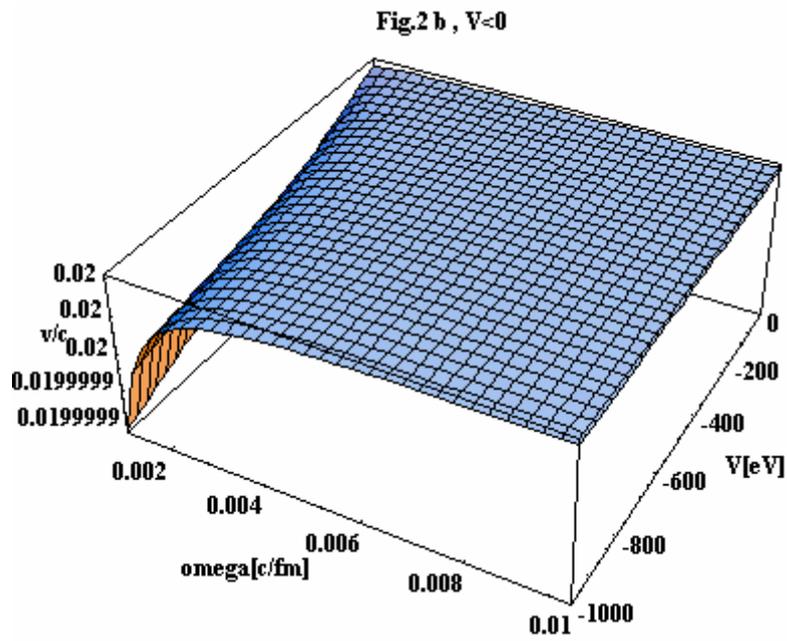

Fig. 2(a) The relative speed of heat propagation for the $\tau = 10^{-17}$ s [2] for $V > 0$, (b) The same for $V < 0$.